\documentclass[one column]{IEEEtran}
\usepackage[latin9]{inputenc}
\usepackage{fancyhdr}
\usepackage{color}
\usepackage{float}
\usepackage{amsmath}
\usepackage{graphicx}
\usepackage[unicode=true,pdfusetitle,
 bookmarks=true,bookmarksnumbered=true,bookmarksopen=true,bookmarksopenlevel=1,
 breaklinks=false,pdfborder={0 0 1},backref=false,colorlinks=false]
 {hyperref}

\makeatletter

\floatstyle{ruled}
\newfloat{algorithm}{tbp}{loa}
\providecommand{\algorithmname}{Algorithm}
\floatname{algorithm}{\protect\algorithmname}

\newcommand{\lyxaddress}[1]{
	\par {\raggedright #1
	\vspace{1.4em}
	\noindent\par}
}

\usepackage{amsfonts}
\usepackage{amsmath}
\usepackage{amsthm}
\usepackage{multirow}

\usepackage{algorithm}
\usepackage{algpseudocode}
\usepackage{setspace}

\usepackage{fancyhdr}

\@ifundefined{showcaptionsetup}{}{%
 \PassOptionsToPackage{caption=false}{subfig}}
\usepackage{subfig}
\makeatother

\begin{document}
\title{Joint Design of Transmit Waveforms and Receive Filters for MIMO Radar
via Manifold Optimization}
\author{Huanyu Zhang and Ziping Zhao\vspace{-20bp}
}
\maketitle

\lyxaddress{\begin{center}
School of Information Science and Technology, ShanghaiTech University,
Shanghai, China\vspace{-20bp}
\par\end{center}}

\lyxaddress{\begin{center}
\{zhanghy4, zhaoziping\}@shanghaitech.edu.cn
\par\end{center}}
\begin{abstract}
The problem of joint design of transmit waveforms and receive filters
is desirable in many application scenarios of multiple-input multiple-output
(MIMO) radar systems. In this paper, the joint design problem is investigated
under the signal-to-interference-plus-noise ratio (SINR) performance
metric, in which case the problem is formulated to maximize the SINR
at the receiver side subject to some practical transmit waveform constraints.
A numerical algorithm is proposed for problem resolution based on
the manifold optimization method, which has been shown to be powerful
and flexible to address nonconvex constrained optimization problems
in many engineering applications. The proposed algorithm is able to
efficiently solve the SINR maximization problem with different waveform
constraints under a unified framework. Numerical experiments show
that the proposed algorithm outperforms the existing benchmarks in
terms of computation efficiency and achieves comparable SINR performance.
\end{abstract}

\vspace{-5bp}

\begin{IEEEkeywords}
MIMO system, SINR maximization, waveform constraints, manifold optimization,
Riemannian gradient descent. 
\end{IEEEkeywords}

\vspace{-10bp}

\section{Introduction}

\label{sec:intro}

Multiple-input multiple-output (MIMO) radar systems have attracted
a lot of attentions due to its flexibility in transmitting different
waveforms through multiple transmit antennas \cite{li_mimo_2009}.
For different application scenarios, the waveforms in a MIMO radar
system can be properly designed to achieve a desired target measured
by a specific performance criterion, which may not be possible in
the classical phased-array radar systems \cite{li_mimo_2007}. Hence,
the intriguing property of waveform diversity has provided the MIMO
radars many appealing features like higher resolution property and
better parameter identifiability property \cite{zhao_joint_2020}.

The problem of joint design of transmit waveforms and receive filters
is desirable in many application scenarios of the MIMO radar systems.
In this paper, we study the joint design problem to maximize the signal-to-interference-plus-noise
ratio (SINR) performance metric at the system receiver side subject
to some practical transmit waveform constraints \cite{wu_transmit_2018}.
The problem is intrinsically non-convex due to the highly non-convex
fractional objective and the non-convex waveform constraints. Since
no analytical solution to the SINR maximization problem can be attained,
many iterative algorithms have been applied to solve it. One of the
classical methods is the sequential optimization algorithm which is
based on the semidefinite relaxation (SDR) with randomization for
rank-1 solution reconstruction \cite{cui_mimo_2014}. Solving an SDR
in each iteration has been argued to have high computational complexity\textcolor{black}{{}
\cite{Luo2010SemidefiniteRO},} which is not applaudable and amenable
to large-scale problems and real-time applications. In order to reduce
the complexity, a widely used method is to resort to the majorization-minimization
(MM) method \cite{hunter_tutorial_2004}. The MM method converts the
original non-convex problem to a series of relatively simpler problems
to be solved in each iteration by choosing a proper upper-bound function.
The MM-based algorithm has been shown to be efficient for the SINR
maximization problem \cite{wu_transmit_2018}. Besides, due to its
flexibility in choosing the upper-bound function, the MM-based algorithm
is able to handle various practical waveform constraints which are
not feasible by SDR method.

Recently, the manifold optimization has shown its advantages in dealing
these non-convex optimization problems for applications in many fields
\cite{hu_brief_2020}. In manifold optimization, amounts of constrained
optimization problems in the Euclidean space can be regarded as unconstrained
optimization problems on the manifolds \cite{absil_optimization_2008}.
Therefore, unconstrained optimization methods (such as the gradient
descent and conjugate gradient) can be implemented on the manifold.
Similar to other engineering fields, manifold optimization methods
have been exploited for problem solving in MIMO radar systems. In
\cite{alhujaili_transmit_2019}, a manifold optimization method called
Riemannian gradient descent (RGD) has been applied for transmit beampattern
synthesis under the unimodular constraint which is modeled as the
complex circle manifold (CMM). However, besides unimodular constraint
there are several other waveform constraints which have practical
applicability with the consideration of hardware configuration. Besides
that, there are few literatures studying the joint design of transmit
waveforms and receive filters for SINR maximization problem. In this
paper, the SINR maximization problem will be studied based on manifold
optimization under multiple waveform constraints, where a unified
projection operator and a unified retraction operator are defined
to help to handle the waveform constraints. Numerical results depict
that our algorithm outperforms the state-of-the-art methods in terms
of computation efficiency and is able to achieve comparable SINR's.

\section{Joint Tx-Rx Design for SINR Maximization}

\label{sec:format}

A MIMO radar system with $N_{t}$ transmit antennas and $N_{r}$ receive
antennas is considered. Each transmit antenna can emit individual
waveform and the $n$-th sample emitted from the $N_{t}$ transmitters
is $\mathbf{s}(n)=[s_{1}(n),\ldots,s_{N_{t}}(n)]^{T}\in\mathbb{C}^{N_{t}}$
with $n=1,\ldots,N$, where $N$ denotes the total number of transmitted
samples. The range-angle position of the target to be tracked is configured
as $(r_{0},\theta_{0})$ and usually we set $r_{0}=0$. Additionally,
$K$ signal-dependent interferers located at $(r_{k},\theta_{k})$
are also taken into account with the range position $r_{k}\in\{0,\ldots,N\}$
and the spatial angle $\theta_{k}\in\{0,\ldots,L\}\times\frac{2\pi}{L+1}$
for $\theta_{k}\neq\theta_{0}$ with $k=1,\dots,K$ and $L$ denoting
the number of discrete azimuth sectors. Therefore, the signals at
the receive antennas can be represented by
\begin{equation}
\setlength{\abovedisplayskip}{2pt}\setlength{\belowdisplayskip}{2pt}\mathbf{x}(n)=\alpha\mathbf{a}_{r}(\theta_{0})\mathbf{a}_{t}(\theta_{0})^{T}\mathbf{s}(n)+\mathbf{d}(n)+\mathbf{v}(n),\label{eq:1}
\end{equation}
for $n=1,\ldots,N$. In \eqref{eq:1}, $\alpha$ is the complex amplitude
of the target with $\mathbb{E}\left[|\alpha|^{2}\right]=\sigma_{0}^{2}$,
and $\mathbf{a}_{r}(\theta)\in\mathbb{C}^{N_{r}}$ and $\mathbf{a}_{t}(\theta)\in\mathbb{C}^{N_{t}}$
are the propagation vector and the steering vector, respectively with
$\mathbf{a}_{t}(\theta)=\frac{1}{\sqrt{N_{t}}}\left[e^{-j\pi0\sin\theta},\ldots,e^{-j\pi(N_{t}-1)\sin\theta}\right]^{T}$
and $\mathbf{a}_{r}(\theta)=\frac{1}{\sqrt{N_{r}}}\left[e^{-j\pi0\sin\theta},\ldots,e^{-j\pi(N_{r}-1)\sin\theta}\right]^{T}$
if the transmit and receive antennas are both assumed to be uniform
linear arrays with half-wavelength separation. The term $\mathbf{d}(n)$
denotes $K$ signal-dependent uncorrelated point-like interferers
as $\mathbf{d}(n)=\sum_{k=1}^{K}\alpha_{k}\mathbf{a}_{r}(\theta_{k})\mathbf{a}_{t}(\theta_{k})^{T}\mathbf{s}(n-r_{k})$,
where $\alpha_{k}$ denotes a complex amplitude with $\mathbb{E}\left[|\alpha_{k}|^{2}\right]=\sigma_{k}^{2}$.
The term $\mathbf{v}(n)\in\mathbb{C}^{N_{t}}$ is a noise term with
covariance $\sigma_{v}^{2}\mathbf{I}_{N_{t}}$.

Let $\mathbf{x}=[\mathbf{x}(1)^{T},\ldots,\mathbf{x}(N)^{T}]^{T}$,
$\mathbf{s}=[\mathbf{s}(1)^{T},\ldots,\mathbf{s}(N)^{T}]^{T}$, and
$\mathbf{v}=[\mathbf{v}(1)^{T},\ldots,\mathbf{v}(N)^{T}]^{T}$. We
get the following compact form as
\begin{equation}
\setlength{\abovedisplayskip}{2pt}\setlength{\belowdisplayskip}{2pt}\begin{array}{c}
\mathbf{x}=\alpha\mathbf{A}(r_{0},\theta_{0})\mathbf{s}+\sum_{k=1}^{K}\alpha_{k}\mathbf{A}(r_{k},\theta_{k})\mathbf{s}+\mathbf{v},\end{array}
\end{equation}
where $\mathbf{A}(r_{k},\theta_{k})=\left[\mathbf{I}_{N}\otimes(\mathbf{a}_{r}(\theta_{k})\mathbf{a}_{t}(\theta_{k})^{T})\right]\mathbf{J}_{r_{k}}$
for $k=0,\cdots,K$ is a Hermitian matrix related to position $r_{k}$
and angle $\theta_{k}$ with a shift matrix $\mathbf{J}_{r_{k}}\in\mathbb{R}^{N_{t}N\times N_{t}N}$
given by
\[
\setlength{\abovedisplayskip}{2pt}\setlength{\belowdisplayskip}{2pt}{\scriptstyle \left[\mathbf{J}_{r_{k}}\right]_{m,n}}=\begin{cases}
1, & m-n=N_{t}r_{k}\\
0, & m-n\neq N_{t}r_{k}
\end{cases}{\scriptstyle =\left[\mathbf{J}_{-r_{k}}^{T}\right]_{m,n}}.
\]
For notational simplicity, we denote $\mathbf{A}(r_{k},\theta_{k})=\mathbf{A}_{k}$
hereafter.

Let $\mathbf{w}\in\mathbb{C}^{N_{r}N}$ be the response receive filters,
the SINR \cite{aubry_knowledge-aided_2013} at the output side can
be calculated as 
\begin{equation}
\setlength{\abovedisplayskip}{2pt}\setlength{\belowdisplayskip}{2pt}\text{SINR}=\frac{\sigma_{0}^{2}\left|\mathbf{w}^{H}\mathbf{A}_{0}\mathbf{s}\right|^{2}}{\mathbf{w}^{H}(\sum_{k=1}^{K}\sigma_{k}^{2}\mathbf{A}_{k}\mathbf{s}\mathbf{s}^{H}\mathbf{A}_{k}^{H})\mathbf{w}+\sigma_{v}^{2}\mathbf{w}^{H}\mathbf{w}}.
\end{equation}

Finally, the joint design of transmit waveforms and receive filters
for SINR maximization (TxRx-SINR) problem is given as
\begin{equation}
\setlength{\abovedisplayskip}{2pt}\setlength{\belowdisplayskip}{2pt}\hspace{-30bp}\begin{aligned} & \underset{\mathbf{s,\;w}}{\text{maximize}} &  & \hspace{-7bp}\frac{\left|\mathbf{w}^{H}\mathbf{A}_{0}\mathbf{s}\right|^{2}}{\mathbf{w}^{H}\hspace{-2bp}\sum_{k=1}^{K}\vartheta_{k}\mathbf{A}_{k}\mathbf{s}\mathbf{s}^{H}\mathbf{A}_{k}^{H}\hspace{-1bp}\mathbf{w}\hspace{-3bp}+\hspace{-3bp}\mathbf{w}^{H}\hspace{-1bp}\mathbf{w}}\\
 & \text{subject to} &  & \hspace{-7bp}\mathbf{s}\in\mathcal{M},
\end{aligned}
\hspace{-6bp}\tag{TxRx-SINR}\hspace{-20bp}\label{eq:prob}
\end{equation}
where $\vartheta_{k}=\sigma_{k}^{2}/\sigma_{v}^{2}>0$, and $\mathcal{M}$
denotes the different considered waveform constraints to be detailed
later.

\section{ALGORITHMIC FRAMEWORK}

\label{sec:pagestyle}

\subsection{Optimization over a manifold\label{subsec:RGD}}

Consider a constrained optimization problem as follows:
\[
\setlength{\abovedisplayskip}{2pt}\setlength{\belowdisplayskip}{2pt}\begin{aligned} & \underset{\mathop{x}}{\text{minimize}} &  & f(x) & \text{subject to} &  & \mathop{x\in\mathcal{R}},\end{aligned}
\]
where $\mathcal{R}$ \textit{is }a constraint set\textit{ }\textit{\emph{treated
as a}}\textit{ }Riemannian\textit{ }\textit{\emph{manifold embedded}}\textit{
}in an Euclidean space $\mathcal{E}\supseteq\mathcal{R}$ equipping
the Riemannian metric \cite{absil_optimization_2008}. By doing this,
optimizing $f(x)$ can be viewed as an unconstrained optimization
problem in the manifold $\mathcal{M}$ rather than a constrained one
with explicit constraint $\mathcal{R}$. Hence, numerous unconstrained
optimization algorithms like the gradient descent \textcolor{black}{\cite{lemarechal2012cauchy}
}can be utilized to handle these manifold optimization problems. 

In this paper, the classical unconstrained optimization method gradient
descent will be implemented for optimization over the Riemannian manifold,
which hence is named as Riemannian gradient descent (RGD) \cite{boumal2020introduction}.
Given an initialization $x^{(0)}$, a sequence $\left\{ x^{(i)}\right\} $
is generated by RGD through iteratively taking two steps until convergence.
The first step is ``descent with projection'' where the gradient
of any smooth extension of the objective function, i.e., $\bar{f}(x)$
with $x\in\mathcal{E}$ is computed as $\nabla\bar{f}(x^{(i)})$,
i.e., the standard gradient in the Euclidean space, then the Riemannian
(manifold) gradient is obtained by projecting $\nabla\bar{f}(x^{(i)})$
onto the tangent space $T_{x^{(i)}}\mathcal{\mathcal{M}}$ by an orthogonal
\textit{\emph{projection at $x^{(i)}$ denoted by}} $\mathsf{Proj}_{x^{(i)}}(\cdot)$,
and finally $\bar{x}^{(i)}$ is obtained by a descent step on $T_{x^{(i)}}\mathcal{\mathcal{M}}$
with the direction $\nabla\bar{f}(x^{(i)})$ and a prespecified stepsize
$\alpha^{(i)}$. Due to the updated $\bar{x}_{k}$ is on $T_{x^{(i)}}\mathcal{\mathcal{M}}$
rather than the manifold $\mathcal{M}$, a \textit{\emph{``retraction''
at $\bar{x}^{(i)}$ denoted by the }}operator $\mathsf{Retr}(\cdot)$
is applied in the second step to map it back to $\mathcal{M}$. To
summarize, the update step of RGD at the $i$-th iteration is
\[
\setlength{\abovedisplayskip}{2pt}\setlength{\belowdisplayskip}{2pt}\begin{cases}
\bar{x}^{\left(i+1\right)}=x^{(i)}-\alpha^{(i)}\mathsf{Proj}_{x^{(i)}}(\nabla\bar{f}(x^{(i)})) & \text{[descent with projection step]}\\
x^{(i+1)}=\mathsf{Retr}(\bar{x}^{(i+1)}) & \text{[retraction step]},
\end{cases}
\]
where stepsize $\alpha^{(i)}$ can be chosen to be constant or according
to a specific stepsize rule like the Armijo back-tracking line search\textcolor{black}{{}
\cite{armijo1966}} for convergence guarantee, and the projection
operator $\mathsf{Proj}_{x^{(i)}}(\cdot)$ and the retraction operator
$\mathsf{Retr}(\cdot)$ vary from manifolds.

\subsection{The projection and retraction in TxRx-SINR\label{subsec:The-proj-and-retr}}

In this section, we consider the projection operators and the retraction
operators w.r.t. different constraints ${\cal M}$'s encountered in
the TxRx-SINR problem \eqref{eq:prob}. Three manifold constraints
are considered which are highly non-convex in the Euclidean space,
namely the constant modulus (CM) constraint $\mathcal{{\cal M}}_{c}=\{\mathbf{s}\,|\,\left\vert s_{n}\right\vert =\frac{1}{\sqrt{NN_{t}}}\}$\textcolor{black}{{}
\cite{he2012waveform}} (including the unimodular constraint, i.e.,
the CCM with ${\normalcolor \left|s_{n}\right|=1}$), the $\epsilon$-uncertainty
constant modulus ($\epsilon$-CM) constraint ${\cal M}_{e}=\{\mathbf{s}\,|\,c_{m}-\epsilon_{1}<\left|s_{n}\right|<c_{m}+\epsilon_{2}\text{ with }0\leq\epsilon_{1}\leq c_{m}\text{ and }0\leq\epsilon_{2}\}$
\cite{zhao_unified_2017}, and the constant modulus and similarity
(CM\&S) constraint ${\cal M}_{s}=\{\mathbf{s}\,|\,\left\vert s_{n}\right\vert =\frac{1}{\sqrt{NN_{t}}},\left\Vert \mathbf{s}-\mathbf{s}_{\text{ref}}\right\Vert _{\infty}\leq\epsilon,\text{ with }0\leq\epsilon\leq\frac{2}{\sqrt{NN_{t}}}\}$
\cite{2009Design}.

\textbf{Projection step.} The projection operator $\mathsf{Proj}_{\mathbf{s}^{(i)}}(\text{\ensuremath{\cdot}})$
at an iterate $\mathbf{s}^{(i)}\in\mathcal{{\cal M}}$ with $\mathcal{{\cal M}}$
taking ${\cal {\cal M}}_{c}$'s or ${\cal {\cal M}}_{s}$'s is the
same and has a closed-form solution. This result is classical in manifold
optimization and can be easily proved by first showing the complex
scalar case and then extending it to to the vector case \cite{boumal2020introduction}.
For any $\mathbf{u}\in\mathbb{C}^{NN_{t}}$, the projection operator
for these two constraints is given by
\begin{equation}
\setlength{\abovedisplayskip}{2pt}\setlength{\belowdisplayskip}{2pt}\mathsf{Proj}_{\mathbf{s}^{\left(i\right)}}^{{\cal {\cal M}}_{c}/{\cal {\cal M}}_{s}}(\mathbf{u})=\mathbf{u}-\mathsf{Re}\left\{ \mathbf{u}^{*}\odot\mathbf{s}^{\left(i\right)}\right\} \odot\mathbf{s}^{\left(i\right)},\label{eq:projection on CCM}
\end{equation}
where $\odot$ denotes the Hadamard product. The $\epsilon$-CM constraint
describes an annulus manifold, the projection operator of which is
given by
\begin{equation}
\setlength{\abovedisplayskip}{2pt}\setlength{\belowdisplayskip}{2pt}\mathsf{Proj}_{\mathbf{s}^{\left(i\right)}}^{\mathcal{{\cal M}}_{e}}(\mathbf{u})=\mathbf{u},\label{eq:projection on epsilon}
\end{equation}

\textbf{Retraction step.} The retraction operators $\mathsf{Retr}(\cdot)$'s
w.r.t. different ${\cal M}$'s can be solved in closed-forms. For
a given $\mathbf{u}\in\mathbb{C}^{NN_{t}}$, a unified retraction
function can be employed to handle all the manifold constraints, which
is given by
\begin{equation}
\setlength{\abovedisplayskip}{2pt}\setlength{\belowdisplayskip}{2pt}\mathsf{Retr}(\mathbf{u})=\arg\min_{\mathbf{s}\in\mathcal{{\cal M}}}\left\Vert \mathbf{s}-\mathbf{u}\right\Vert _{2},\label{eq:uni-ret}
\end{equation}
where specifically the solution w.r.t. $\mathcal{M}_{c}$ is given
by $\mathsf{Retr}(\mathbf{u})=\mathbf{u}\odot\bigl(\sqrt{NN_{t}}\text{\ensuremath{\left|\mathbf{u}\right|}}\bigr)^{-1}$
with $\left|\cdot\right|$ and $(\cdot)^{-1}$ applied element-wisely,
w.r.t. $\mathcal{{\cal M}}_{s}$ is given in \cite{cui_space-time_2017},
and w.r.t. ${\cal M}_{e}$ is given in \cite{zhao_unified_2017}.

\section{Solving the TxRx-SINR problem via RGD}

Now we are ready to derive the RGD algorithm for problem (\ref{eq:prob}).
Noting that problem (\ref{eq:prob}) is invariant to a scaling in
$\mathbf{w}$, for a fixed $\mathbf{s}$ it can be transformed to
be a convex problem as
\begin{equation}
\setlength{\abovedisplayskip}{2pt}\setlength{\belowdisplayskip}{2pt}\begin{aligned} & \underset{\mathbf{w}}{\text{minimize}} &  & \hspace{-5bp}\begin{array}{c}
\mathbf{w}^{H}\left[\sum_{k=1}^{K}\vartheta_{k}\mathbf{A}_{k}\mathbf{s}\mathbf{s}^{H}\mathbf{A}_{k}^{H}+\mathbf{I}\right]\mathbf{w}\end{array}\\
 & \text{subject to} &  & \mathbf{w}^{H}\mathbf{A}_{0}\mathbf{s}=1,
\end{aligned}
\tag{Rx Prob.}\label{eq:Rx Prob}
\end{equation}
to which a closed-form solution for $\mathbf{w}$ is obtained by \textcolor{black}{\cite{boyd2004convex}}
\begin{equation}
\setlength{\abovedisplayskip}{2pt}\setlength{\belowdisplayskip}{2pt}\hspace{-8bp}\mathbf{w}^{\star}\hspace{-3bp}=\hspace{-3bp}\frac{\left[\sum_{k=1}^{K}\vartheta_{k}\mathbf{A}_{k}\mathbf{s}\mathbf{s}^{H}\mathbf{A}_{k}^{H}+\mathbf{I}\right]^{-1}\hspace{-3bp}\mathbf{A}_{0}\mathbf{s}}{\mathbf{s}^{H}\mathbf{A}_{0}^{H}\hspace{-3bp}\left[\sum_{k=1}^{K}\vartheta_{k}\mathbf{A}_{k}\mathbf{s}\mathbf{s}^{H}\mathbf{A}_{k}^{H}+\mathbf{I}\right]^{-1}\hspace{-3bp}\mathbf{A}_{0}\mathbf{s}}.\hspace{-5bp}\tag{Optim. Rx}\label{eq:15}
\end{equation}

Substituting \eqref{eq:15} into the original problem \ref{eq:prob},
we get the subproblem for the transmit waveforms as
\begin{equation}
\setlength{\abovedisplayskip}{2pt}\setlength{\belowdisplayskip}{2pt}\hspace{-19bp}\begin{aligned} & \underset{\mathbf{s}}{\text{minimize}} &  & \hspace{-8bp}-\hspace{-2bp}\mathbf{s}^{H}\hspace{-3bp}\mathbf{A}_{0}^{H}\hspace{-3bp}\left[\sum_{k=1}^{K}\vartheta_{k}\mathbf{A}_{k}\mathbf{s}\mathbf{s}^{H}\mathbf{A}_{k}^{H}\hspace{-2bp}+\hspace{-2bp}\mathbf{I}\right]^{-1}\hspace{-8bp}\mathbf{A}_{0}\mathbf{s}\\
 & \text{subject to} &  & \hspace{-5bp}\mathbf{s}\in\mathcal{M},
\end{aligned}
\hspace{-15bp}\tag{Tx Prob.}\label{eq:Tx prob}
\end{equation}
where the waveform constraints $\mathcal{M}$ can take different forms
as discussed in Sec. \ref{subsec:The-proj-and-retr}. 

To solve the original problem \ref{eq:prob}, it suffices to solve
the problem \eqref{eq:Tx prob} for $\mathbf{s}$ and then obtain
$\mathbf{w}$ by \eqref{eq:15}. In this paper, we propose to solve
\eqref{eq:Tx prob} via the RGD method in Sec. \ref{subsec:RGD}.
For convergence concern, the objective of problem (\ref{eq:Tx prob})
will be augmented with a constant term $\gamma\mathbf{s}^{H}\mathbf{s}$
w.l.o.g. ($\gamma$ is a prescribed constant, the choice of which
guarantees the monotonicity of ``projection'' step in RGD and is discussed
in an online supplementary material\textcolor{blue}{{} }\cite{monotonicity}
due to space limitation) to control the monotonicity of the retraction
operator $\mathsf{Retr}(\cdot)$. Then we define the ``augmented''
objective function for problem \eqref{eq:Tx prob} as
\[
\setlength{\abovedisplayskip}{2pt}\setlength{\belowdisplayskip}{2pt}g(\mathbf{s})=-\mathbf{s}^{H}(\mathbf{A}_{0}^{H}\hspace{-3bp}\left[\sum_{k=1}^{K}\vartheta_{k}\mathbf{A}_{k}\mathbf{s}\mathbf{s}^{H}\mathbf{A}_{k}^{H}+\mathbf{I}\right]^{-1}\hspace{-8bp}\mathbf{A}_{0})\mathbf{s}+\gamma\mathbf{s}^{H}\mathbf{s}.
\]
The gradient of a smooth extension of the objective function denoted
by $\bar{g}(\mathbf{s})$ (extending $g(\mathbf{s})$ to the Euclidean
domain) is given by 
\begin{equation}
\hspace{-10bp}\begin{aligned}\nabla\bar{g}(\mathbf{s})= & -2\left(\mathbf{A}_{0}^{H}\left(\sum_{k=1}^{K}\vartheta_{k}\mathbf{A}_{k}\mathbf{s}\mathbf{s}^{H}\mathbf{A}_{k}^{H}+\mathbf{I}\right)^{-1}\mathbf{A}_{0}\mathbf{s}\right)+2\gamma\mathbf{s}-\left(\mathbf{s}^{H}\frac{\partial}{\partial\mathbf{s}}\left(\mathbf{A}_{0}^{H}\left(\sum_{k=1}^{K}\vartheta_{k}\mathbf{A}_{k}\mathbf{s}\mathbf{s}^{H}\mathbf{A}_{k}^{H}+\mathbf{I}\right)^{-1}\mathbf{A}_{0}\right)\right)\mathbf{s}.\\
= & -2\left(\mathbf{A}_{0}^{H}\left(\sum_{k=1}^{K}\vartheta_{k}\mathbf{A}_{k}\mathbf{s}\mathbf{s}^{H}\mathbf{A}_{k}^{H}+\mathbf{I}\right)^{-1}\mathbf{A}_{0}\mathbf{s}\right)+2\gamma\mathbf{s}-\left(\mathbf{1}_{NN_{t}}\otimes\mathbf{s}^{H}\mathbf{A}_{0}^{H}\left(\sum_{k=1}^{K}\vartheta_{k}\mathbf{A}_{k}\mathbf{s}\mathbf{s}^{H}\mathbf{A}_{k}^{H}+\mathbf{I}\right)^{-1}\right)\\
 & \qquad\qquad\qquad\qquad\times\sum_{k=1}^{K}\left(\vartheta_{k}\mathbf{I}_{NNt}\otimes\mathbf{A}_{k}\left[\frac{\partial\mathbf{s}\mathbf{s}^{H}}{\partial s_{1}}\:\ldots\:\frac{\partial\mathbf{s}\mathbf{s}^{H}}{\partial s_{NN_{t}}}\right]^{T}\mathbf{A}_{k}^{H}\right)\left(\sum_{k=1}^{K}\vartheta_{k}\mathbf{A}_{k}\mathbf{s}\mathbf{s}^{H}\mathbf{A}_{k}^{H}+\mathbf{I}\right)^{-1}\mathbf{A}_{0}\mathbf{s},
\end{aligned}
\label{eq:longlong}
\end{equation}
 where $\otimes$ denotes the Kronecker product and $\frac{\partial\mathbf{s}\mathbf{s}^{H}}{\partial s_{n}}$
($n=1,\ldots,NN_{t}$) is a matrix calculated by $\frac{\partial\mathbf{s}\mathbf{s}^{H}}{\partial s_{n}}=\left[\mathbf{s},\;\mathbf{0}\;,\ldots,\;\mathbf{0}\right]\mathbf{J}_{r_{k}/N_{t}}+\mathbf{J}_{r_{k}/N_{t}}^{T}\left[\mathbf{s},\;\mathbf{0}\;,\ldots,\;\mathbf{0}\right]^{H}$
with the shift matrix $\mathbf{J}_{r_{k}/N_{t}}\in\mathbb{R}^{N_{t}N\times N_{t}N}$
defined in Sec. \ref{sec:format}.

Finally, the proposed RGD algorithm for TxRx-SINR is summarized in
Algorithm\textcolor{black}{{} \ref{alg:Solving-TxRx-SINR-problem}}\footnote{A well-chosen stepsize $\alpha^{(i)}$ is to ensure the decrease of
the objective function $g\left(\cdot\right)$ in the ``descent with
projection'' step of RGD.}.

{\footnotesize{}}
\begin{algorithm}[t]
{\footnotesize{}\caption{Solving TxRx-SINR problem via RGD\label{alg:Solving-TxRx-SINR-problem}}
}{\footnotesize\par}

{\small{}$\mathbf{Initialize}$: $i=0$, $\mathbf{s}^{(0)}$, $\alpha$,
$\beta$, $\tau$, $\sigma$}{\small\par}

{\small{}$\mathbf{While}$ not converge $\mathbf{do}$}{\small\par}

{\small{}$\,\,$1.$\hspace{1em}$}compute$\nabla\bar{g}(\mathbf{s}^{(i)})$
according to \eqref{eq:longlong}

{\small{}$\,\,$2.$\hspace{1em}$}compute$\mathsf{Proj}_{\mathbf{s}_{k}}(\nabla g(\mathbf{s}^{(i)}))$
according to \eqref{eq:projection on CCM} or \eqref{eq:projection on epsilon}

{\small{}$\,\,$3-1. }{[}$\textbf{constant stepsize}${]} $\alpha^{(i)}=\alpha$

{\small{}$\,\,$3-2. }{[}$\textbf{Armijo back-tracking line search }$\cite{armijo1966}{]}
$\alpha^{(i)}=\tau\beta^{m}$

$\qquad$with $m$ the smallest non-negative integer such that\vspace{-7bp}
\[
g(\mathbf{s}^{(i)})-g(\mathbf{s}^{(i)}-\tau\beta^{m}\nabla g(\mathbf{s}^{(i)}))\geq\sigma\tau\beta^{m}\|\mathsf{Proj}_{s^{(i)}}(\nabla g(\mathbf{s}^{(i)}))\|_{2}^{2}
\]
\vspace{-17bp}

{\small{}$\,\,$4. $\hspace{1em}$}$\bar{\mathbf{s}}^{(i+1)}=\mathbf{s}^{(i)}-\alpha^{(i)}\mathsf{Proj}_{s^{(i)}}(\nabla g(\mathbf{s}^{(i)}))$

{\small{}$\,\,$5. $\hspace{1em}$}$\mathbf{s}^{(i+1)}=\mathsf{Retr}(\mathbf{s}^{(i)})$
according to \eqref{eq:uni-ret}

{\small{}$\,\,$6. $\hspace{1em}$}$i=i+1$

{\small{}$\mathbf{end}$ $\mathbf{while}$}{\small\par}

compute $\mathbf{w}$ according to \eqref{eq:15}
\end{algorithm}
{\footnotesize\par}

\section{Numerical Experiments}

\label{sec:typestyle}

\begin{figure}
\begin{centering}
\subfloat[\label{fig:CMC_runtime}CM constraint]{\begin{centering}
\textsf{\includegraphics[width=0.4\columnwidth,height=0.34\columnwidth]{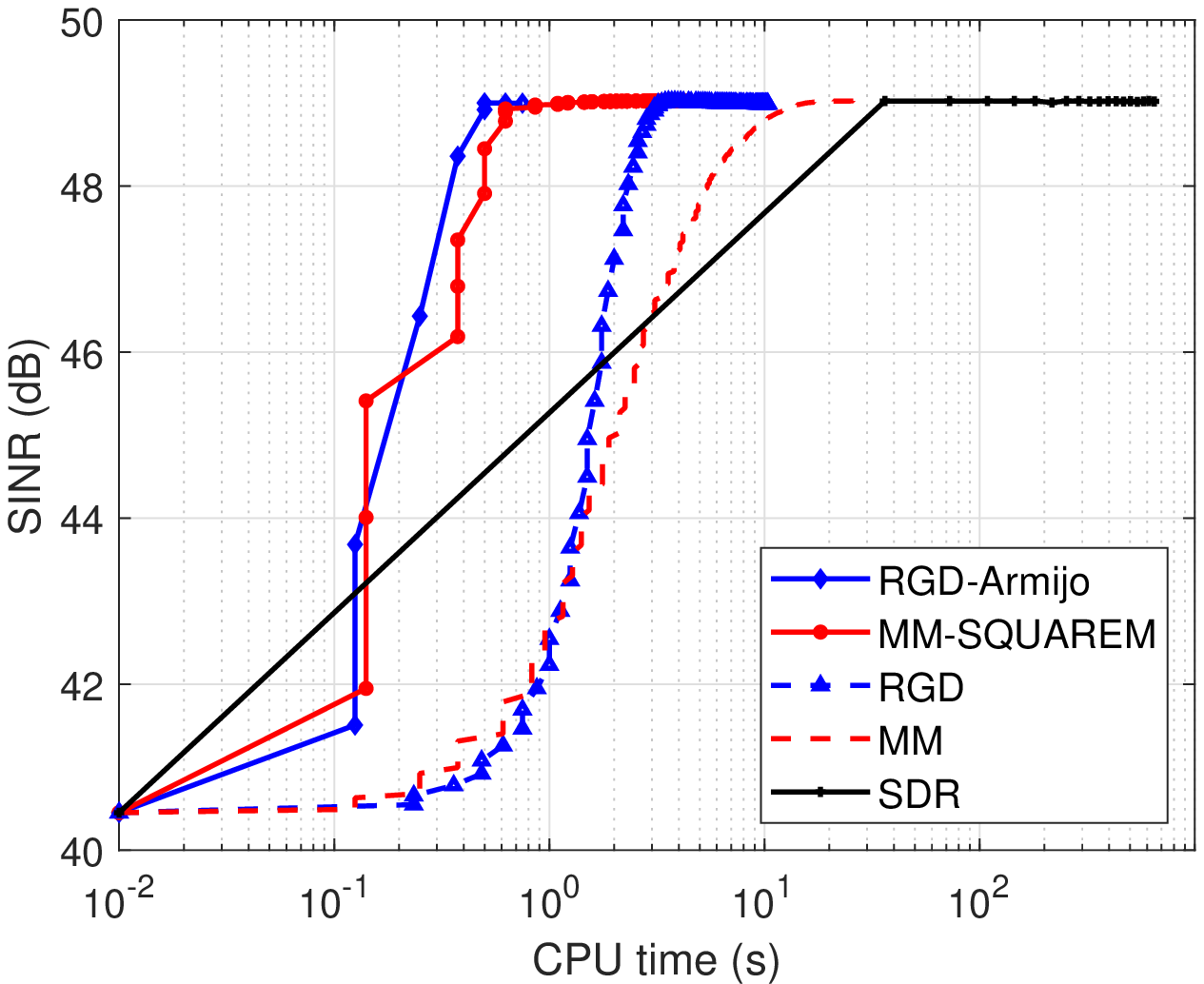}}
\par\end{centering}
}\hspace{5bp}\subfloat[\label{fig:CMC_runtime-1}CM constraint]{\begin{centering}
\textsf{\includegraphics[width=0.4\columnwidth,height=0.34\columnwidth]{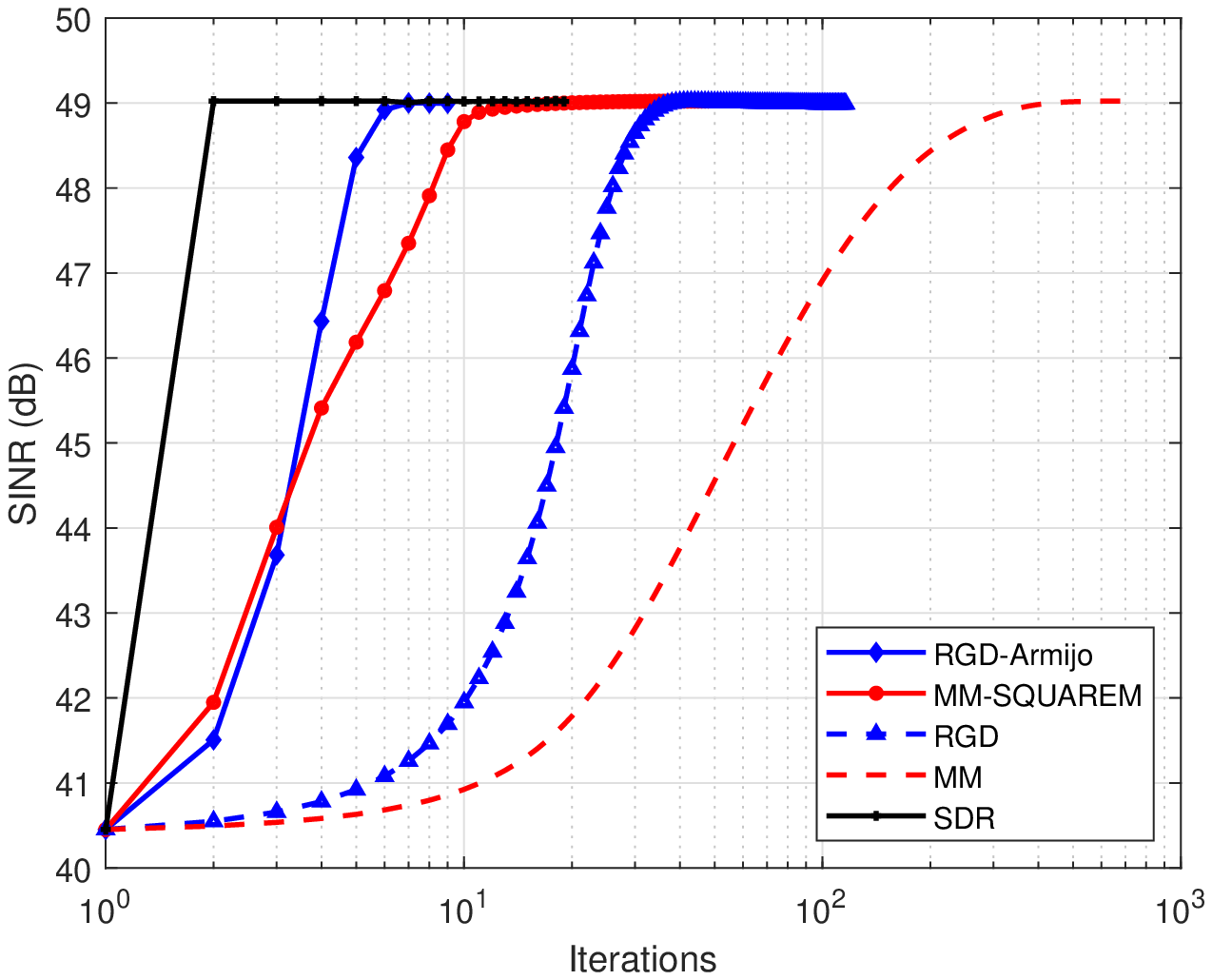}}
\par\end{centering}
}
\par\end{centering}
\begin{centering}
\subfloat[\label{fig:Similar-runtime}CM\&S constraint]{\begin{centering}
\includegraphics[width=0.4\columnwidth,height=0.34\columnwidth]{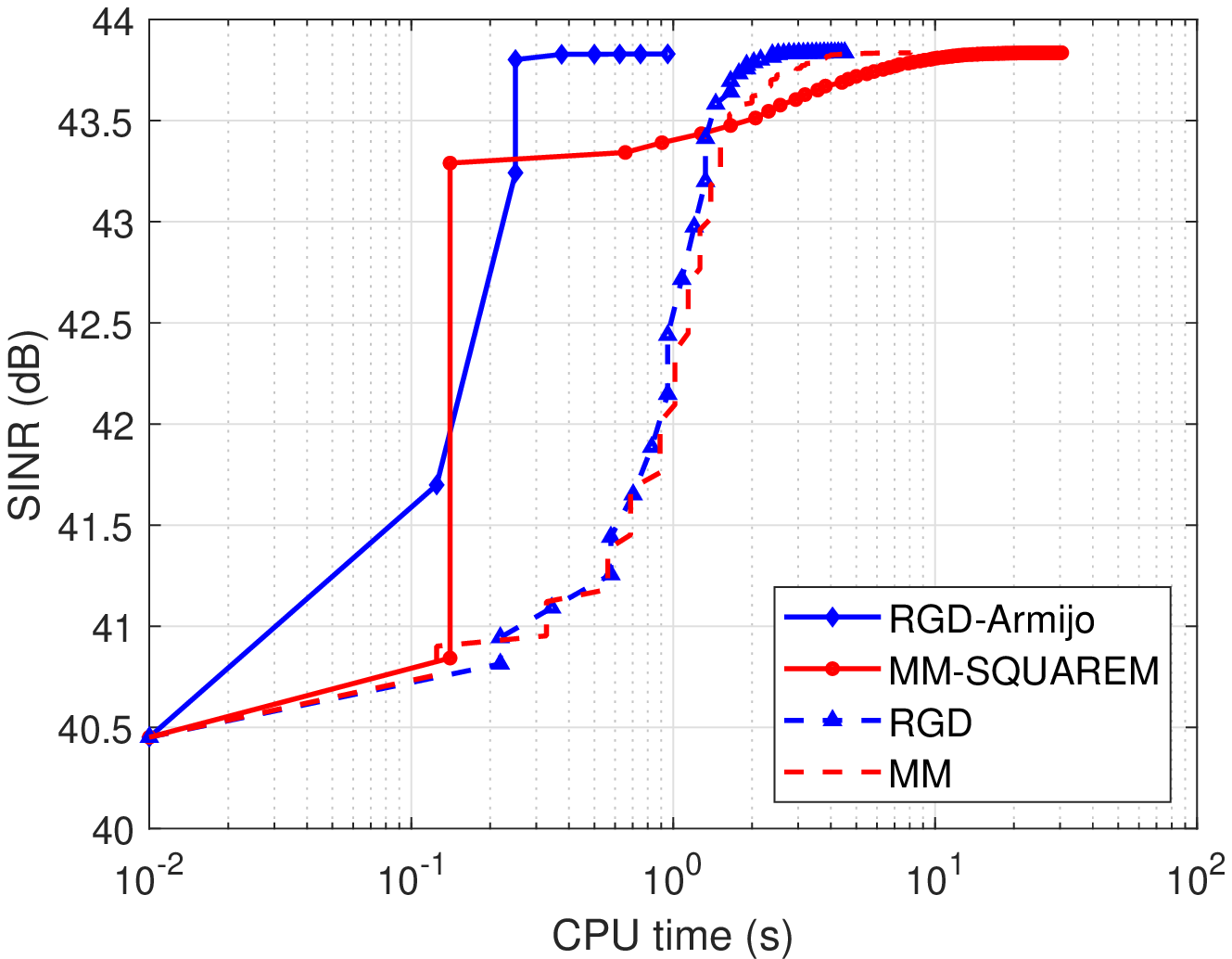}
\par\end{centering}
}\hspace{5bp}\subfloat[\label{fig:Similar-runtime-1}CM\&S constraint]{\begin{centering}
\includegraphics[width=0.4\columnwidth,height=0.34\columnwidth]{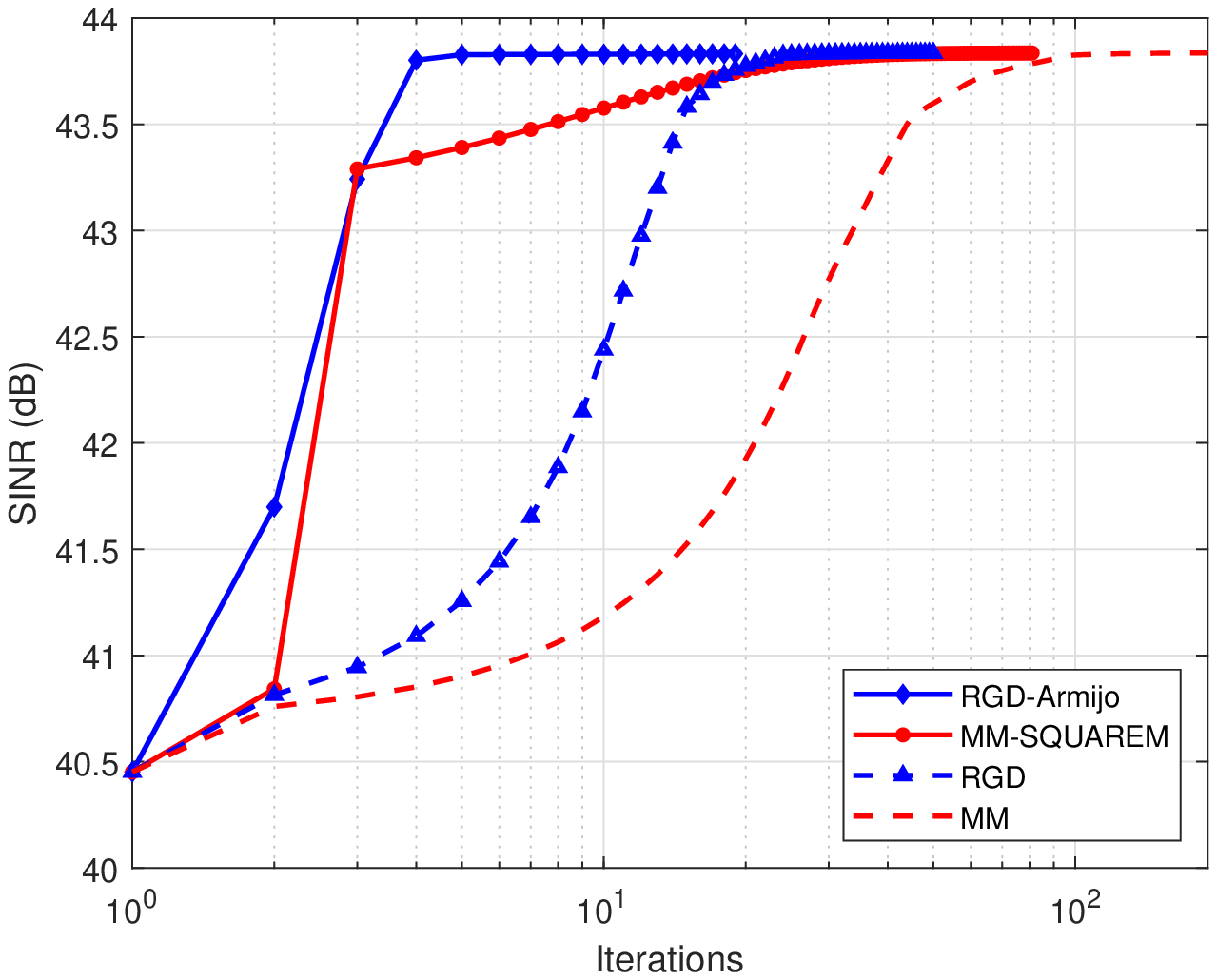}
\par\end{centering}
}
\par\end{centering}
\begin{centering}
\subfloat[\label{fig:e_runtime}$\epsilon$-CM constraint]{\begin{centering}
\includegraphics[width=0.4\columnwidth,height=0.34\columnwidth]{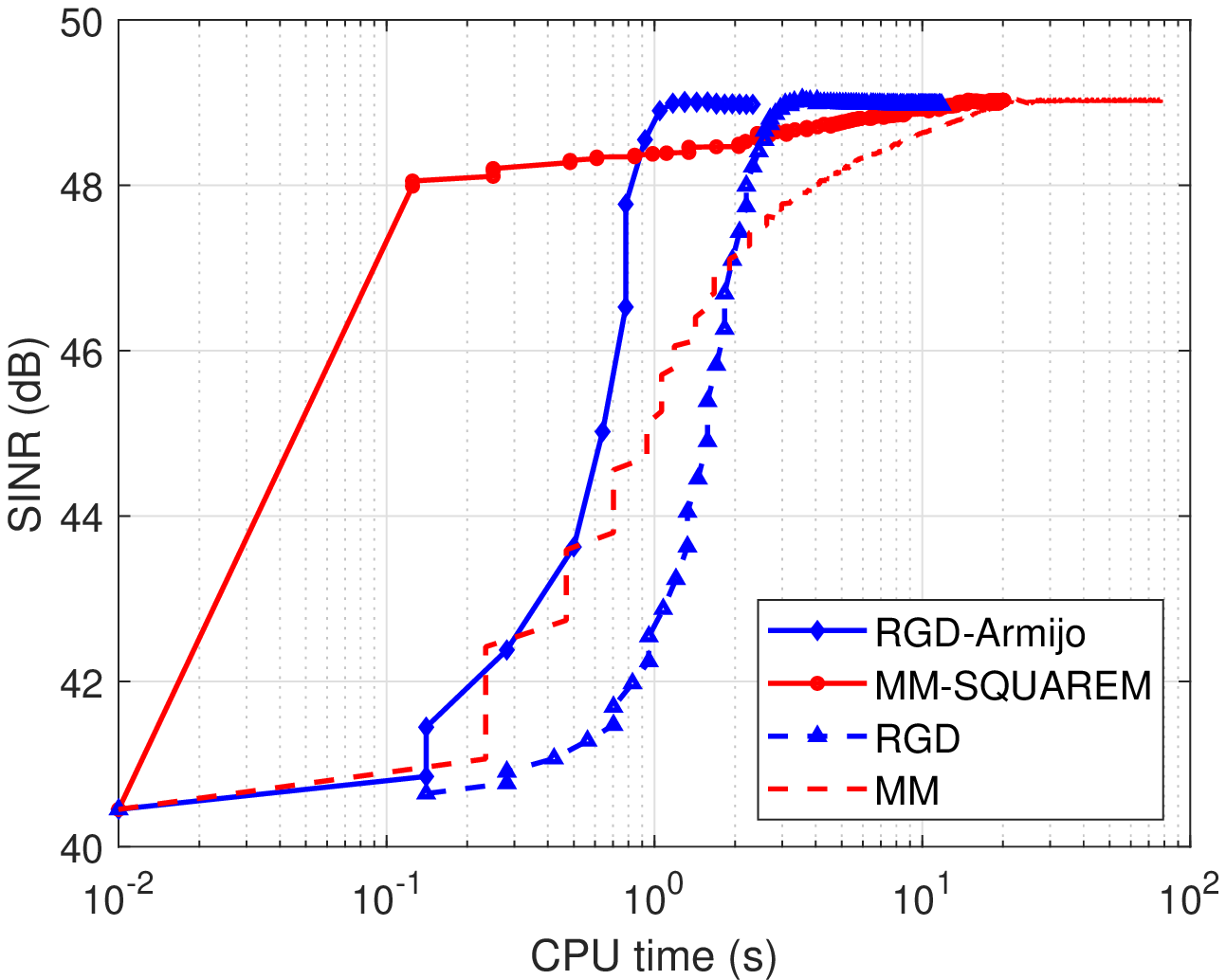}
\par\end{centering}
}\hspace{5bp}\subfloat[\label{fig:e_runtime-2}$\epsilon$-CM constraint]{\begin{centering}
\includegraphics[width=0.4\columnwidth,height=0.34\columnwidth]{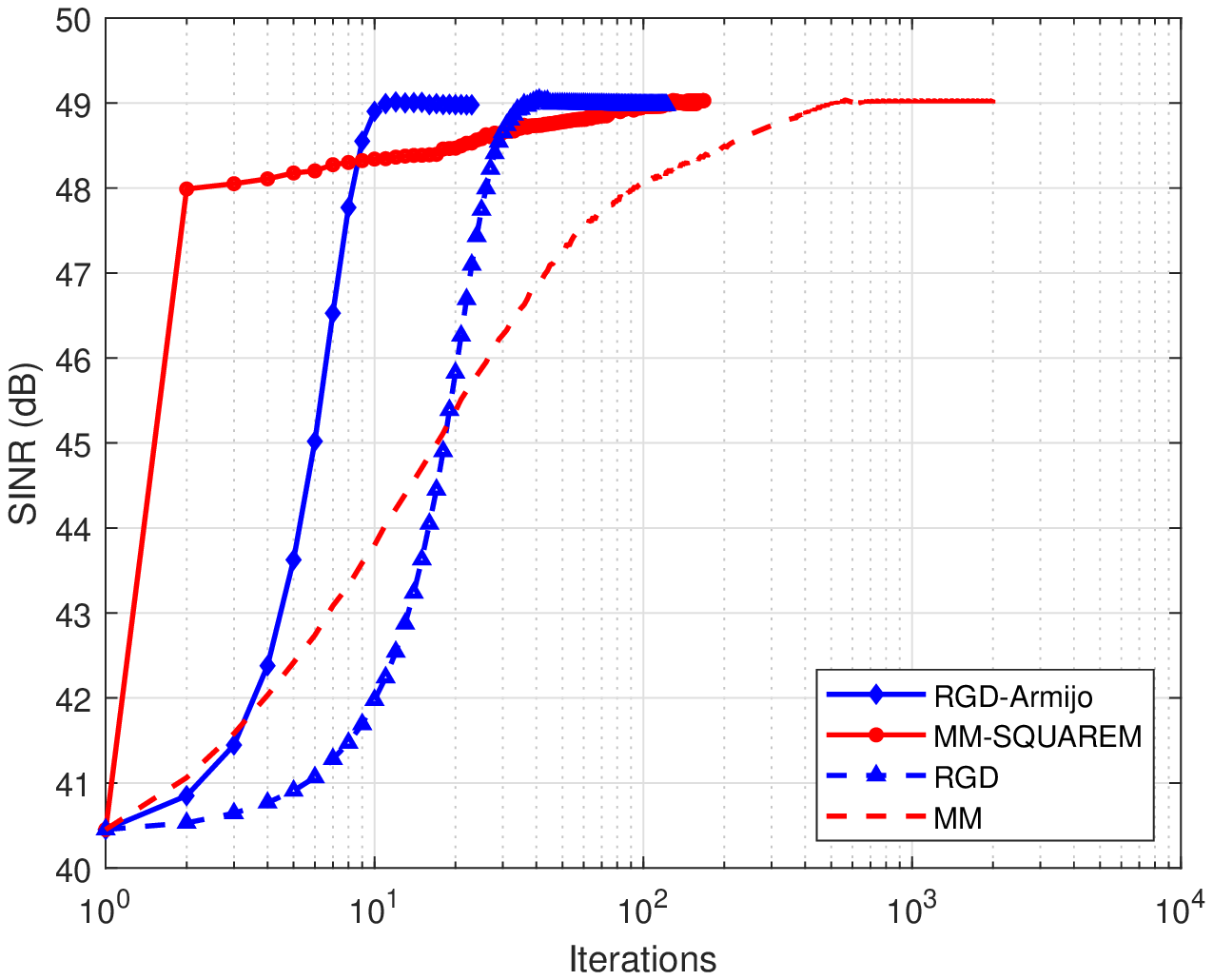}
\par\end{centering}
}
\par\end{centering}
\caption{Convergence rate comparisons.}
\end{figure}

In this section, we compare the performance of our proposed RGD algorithm
with the state-of-the-art methods in the literatures. The simulation
is conducted on the MATLAB2019b platform under a PC machine with an
Intel i7--10700 CPU and 16GB RAM. For the MIMO radar system settings,
the range-angle position of the target to be tracked is configured
as $(0,15^{\circ})$, the power of which is $|\alpha_{0}|^{2}=30$dB.
Three fixed interferers are located at the range-angle positions $(0,-50^{\circ})$,
$(1,-10^{\circ})$, and $(2,40^{\circ})$, respectively. The power
of each interferer is $|\alpha_{j}|^{2}=20$dB for $j=1,2,3$. The
variance of the noise is $\sigma_{v}^{2}=0$dB. The orthogonal linear
frequency modulation (LFM) waveforms are set as the initial and also
the reference waveforms in the CM\&S constraint. The space-time LFM
waveform matrix is

\begin{equation}
\setlength{\abovedisplayskip}{1pt}\setlength{\belowdisplayskip}{1pt}\mathbf{S}^{\left(0\right)}(k,n)=\frac{e^{j2\pi k(n-1)/N}e^{j\pi(n-1)^{2}/N}}{\sqrt{NN_{t}}},
\end{equation}
where $k=1,\cdots,N_{t}$ and $n=1,\cdots N$, based on which we obtain
the initialization iteration $\mathbf{s}^{(0)}=\mathrm{vec}\left(\mathbf{S}^{(0)}\right)$. 

We first compare the performance of the TxRx-SINR problem with CM
constraint between the two proposed algorithms, i.e., the RGD algorithm
and the RGD with Armijo rule denoted as RGD-Armijo (for parameters
in the Armijo back-tracking rule, we have set $\sigma=1$, $\beta=0.85$,
and $\tau=0.4$) with the benchmark methods, namely SDR, MM, and MM
with SQUAREM acceleration denoted as MM-SQUAREM under three different
waveform constraints. The collocated MIMO radar parameters are chosen
as $N_{t}=10,N_{r}=10,$ and $N=8$. In Fig \ref{fig:CMC_runtime}
and \ref{fig:CMC_runtime-1}, it can be shown that all methods converge
to the same SINR. As expected, SDR is the most time-consuming one
within the fewest iterations. RGD-Armijo converges faster than RGD
and both of them perform better than MM. MM-SQUAREM is much faster
than MM due to the acceleration scheme, but is still slower than RGD-Armijo.
Similar convergence results are observed for the TxRx-SINR problem
with other constraints, namely the CM\&S constraint (similarity parameter
$\epsilon=1/\sqrt{N_{t}N}$), and the $\epsilon$-CM constraint, which
are shown in Fig \ref{fig:Similar-runtime}, Fig \ref{fig:Similar-runtime-1},
Fig \ref{fig:e_runtime-2} and Fig \ref{fig:e_runtime}, respectively.

\begin{figure}[t]
\begin{centering}
\includegraphics[width=0.7\columnwidth]{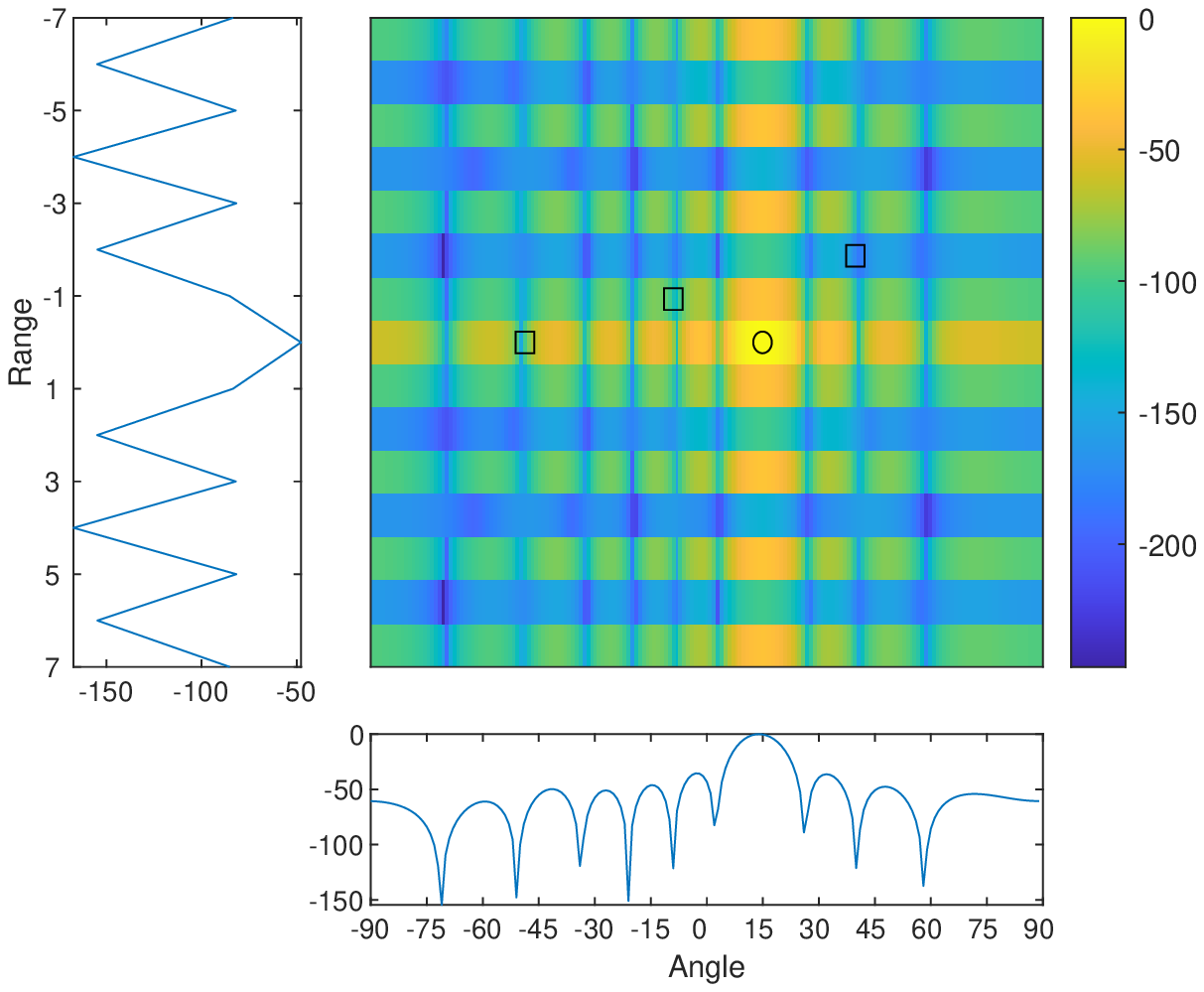}
\par\end{centering}
\caption{\label{fig:e_runtime-1}Range-angle ambiguity function for CM constraint.}
\end{figure}

To further test the scalability of the proposed algorithms, cases
with different ($N$, $N_{r}$, $N_{t}$) are evaluated with comparisons
to the benchmark methods under the CM constraint. The runtimes are
reported in Table \ref{tab:runtime}. To evaluate the performance
of the MIMO ambiguity function shaping via the designed waveforms
from our proposed RGD algorithm. We plot the ambiguity function (the
expression of ambiguity function is chosen as in \cite{rabaste_signal_2013})
in Fig. \ref{fig:e_runtime-1}, which can capture the inherent resolution
properties of the MIMO radar systems \cite{san_antonio_mimo_2007}.
In Fig. \ref{fig:e_runtime-1}, it can be observed that the ambiguity
function resembles a thumbtack, the maximum value of which is located
at $(0,15^{\circ})$ marked by a circle. For the interferers, their
locations are marked by rectangles. Values of interferers in the ambiguity
function are relatively small. Fig. \ref{fig:e_runtime-1} also provides
the angle slice at the range $r=0$ and the range slice at the angle
$\theta=15^{\circ}$. It can be observed that there are cliffs at
angle $\theta=40^{\circ},-10^{\circ}$, and $-50^{\circ}$ with range
$r=2$.

\begin{table}[t]
\caption{Runtime comparisons under CM constraint. \label{tab:runtime}}

\begin{center}
\scalebox{1.1}{
\setlength{\tabcolsep}{2.5mm}{ 
\begin{tabular}{cccccc}  
\hline 
  \multirow{2}{*}{Algorithm}&\multicolumn{4}{c}{($N_r$, $N_t$, $N$)}&\\ \cline{2-6}
                       & {(4,4,4)}&     {(10,10,4)}&{(10,10,8)}&     {(15,15,8)}&     {(10,10,30)}\\  
\hline  
               {RGD-Armijo}&   {0.0197sec.}&        {0.3729sec. }& {0.7536sec. }&        {4.3909sec. }& {74.3386sec. }\\ 
               {MM-SQUAREM}& {0.7969sec. }&       {1.2813sec. }& {2.2971sec. }&        {29.7969sec. }& {518.7628sec. }\\ 
               {RGD}&   {0.7031sec. }&        {3.7134sec. }& {10.3421sec. }&        {173.4153sec. }& {1026.127sec. }\\
			   {MM}& {1.875sec. }&       {4.0156sec. }& {20.3203sec. }&        {207.6719sec. }& {1231.6143sec. }\\ 
               {SDR}&{35.1021sec. }&         {31.4375sec. }& {431.7156sec. }&        {620.8147sec. }& {2029.9058sec. }\\ 
\hline  
\end{tabular}}}
\end{center}
\end{table}
\vspace{10bp}

\section{Conclusions}

\label{sec:majhead}

In this paper, we have considered the SINR maximization problem in
MIMO radar subject to multiple practical waveform constraints by jointly
designing the transmit waveforms and receive filters. A manifold optimization
algorithm called RGD is proposed for problem resolution. Numerical
results validate the superiority of the proposed algorithms. It should
be mentioned that we note there is an independent work on manifold
optimization for the SINR maximization problem available recently
\cite{li_riemannian_2020}, which considers a different system setting
(the airborne MIMO-STAP radar setting) and only focuses on the CM
constraint. Besides that, the RGD algorithm proposed in our paper
is leveraging the variable reduction technique in the alternating
minimization which is different from the technique used in that paper.
\pagebreak{}

\label{sec:refs}

\bibliographystyle{plain}
\bibliography{Design_of_Transmit_Waveforms_and_Receive_Filters_via_Manifold_Optimization}

\end{document}